\documentclass[11pt,onecolumn,a4paper]{article}

\usepackage{amsmath}
\usepackage{amsfonts}
\usepackage{amssymb}
\usepackage{amsthm}

\usepackage{natbib}

\usepackage[latin1]{inputenc}
\usepackage{graphicx,color}
\usepackage{url}
\usepackage{hyperref}
\usepackage{latexsym,enumerate}
\usepackage{booktabs}
\usepackage{framed}
\usepackage{subfigure}
\usepackage{multirow} 
\usepackage{color}
\usepackage{colortbl}
\definecolor{gray1}{rgb}{0.8,0.8,0.8}	
\definecolor{gray2}{rgb}{0.95,0.95,0.95}

\newcommand{\CC}{{\mathbb C}}
\newcommand{\RE}{\,{\rm Re}}
\newcommand{\IM}{\,{\rm Im}}

\renewcommand{\eqref}[1]{Eq. (\ref{#1})}

\newcommand\commentout[1]{}

\begin{document}

\title{
M\"obius Moduli for Fingerprint Orientation Fields 
}


\author{Christina Imdahl\thanks{
K\"uhne Logistics University Hamburg,
Gro{\ss}er Grasbrook 17, 20457 Hamburg, Germany.
Email: christina.imdahl@the-klu.org}%
\and Carsten Gottschlich\thanks{Institute for Mathematical Stochastics,
University of G\"ottingen,
Goldschmidtstr. 7, 37077 G\"ottingen, Germany.
Email: gottschlich@math.uni-goettingen.de}%
\and Stephan Huckemann\thanks{Felix-Bernstein-Institute 
for Mathematical Statistics in the Biosciences,
University of G\"ottingen,
Goldschmidtstr. 7, 37077 G\"ottingen, Germany.
Email: huckeman@math.uni-goettingen.de}
\and Ken'ichi Ohshika\thanks{Department of Mathematics, Graduate School of Science, Osaka University, Toyonaka, Osaka 560-0043, Japan.
Email: ohshika@math.sci.osaka-u.ac.jp}
}

\maketitle
\tableofcontents

\begin{abstract}


We propose a novel fingerprint descriptor, namely M\"obius moduli, measuring local deviation of orientation fields (OF) of fingerprints from conformal fields, and we propose a method to robustly measure them, based on tetraquadrilaterals to approximate a conformal modulus locally with one due to a M\"obius transformation. Conformal fields arise by the approximation of fingerprint OFs given by zero pole models, which are determined by the singular points and a rotation. This approximation is very coarse, e.g. for fingerprints with no singular points (arch type), the zero-pole model's OF has parallel lines. Quadratic differential (QD) models, which are obtained from zero-pole models by adding suitable singularities outside the observation window, approximate real fingerprints much better. For example, for arch type fingerprints, parallel lines along the distal joint change slowly into circular lines around the nail furrow. Still, QD models are not fully realistic because, for example along the central axis of arch type fingerprints, ridge line curvatures usually first increase and then decrease again. It is impossible to model this with QDs, which, due to complex analyticity, also produce conformal fields only. In fact, as one of many applications of the new descriptor, we show, using histograms of curvature and conformality index (log of the absolute value of the M\"obius modulus), that local deviation from conformality in fingerprints occurs systematically at high curvature which is not reflected by state of the art fingerprint models as are used, for instance, in the well known synthetic fingerprint generation tool SFinGe and these differences robustely discriminate real prints from SFinGe's synthetic prints.

\end{abstract}

\section*{Keywords}

Fingerprint recognition, orientation field modeling, M\"obius transformation, conformal modulus,
quadratic differentials, zero-pole model, Riemann mapping theorem

\section*{MSC Class}
Primary: 62P10\\
Secondary: 30C62

\section{Introduction}

Fingerprints are usually described by features at several levels. At the first level there is the orientation field, at the second there are minutiae and ridge frequencies and at the third there are pores etc., for an overview see for instance \cite{MaltoniMaioJainPrabhakar2009}.
In this paper we introduce a new feature at the second level which has previously not been studied. It is a local feature of the orientation field and we call it the \emph{M\"obius modulus}. The log of its absolute value is the \emph{conformality index}. These terms are motivated from the theory of complex functions and in particular from the Riemann mapping theorem. 

Informally, our new feature can be understood as follows. Fix two points $p_1$ and $p_2$ on a common fingerprint ridge, uniformly heat up this ridge segment and follow the diffusion of heat orthogonal to the orientation field. If heat level curves agree with neighboring ridge lines (as in Figure \ref{Riemann-mapping-thm:fig}), we say that the field is locally \emph{conformal}, and indeed, it turns out that fingerprints are locally conformal to large extent. If the ridge lines no longer agree with heat level curves (as in Figure \ref{deviation-conformality:fig}), we say that the field is locally \emph{non-conformal}. In this paper we make these concepts precise and propose a method to locally measure the degree of conformality of fingerprint orientation fields outside neighborhoods of the singular points. At this point we note that in view of application and audience, we resort to the lean notion of \emph{conformal} and \emph{non-conformal} fields, where in the language of Riemann surfaces (e.g. \cite{Strebel1984}) we refer to a conformal structure \emph{equivalent} or \emph{not equivalent} to that induced by the conformal structure of the Riemann sphere.

This new  feature, the conformality index, has several applications. In this paper we show that curvature combined with this new feature is highly discriminatory for distinguishing real fingerprints from fingerprints synthetically generated  by 
SFinGe\footnote{http://biolab.csr.unibo.it/research.asp?organize=Activities\&select=\&selObj=12\&\\pathSubj=111\%7C\%7C12\&}.

There are other applications of M\"obius moduli and the conformality index of which we mention the following. In \cite{HuckemannHotzMunk2008} we have introduced quadratic differentials as global models for orientation fields of fingerprints. They are generalizations of zero pole models from \cite{SherlockMonro1993model}, and such models yield conformal fields, by their very definition via meromorphic functions. Since orientation fields of real fingerprints also feature non-conformality, in order to obtain an asymptotically perfectly fitting orientation field, in the low parameter representation of \cite{GottschlichTamsHuckemann2017}, one may precisely feed in the loci of highest deviation from conformality. This gives a natural low dimensional feature vector, which can be used for fast indexing. As another application, e.g. to latent fingerprint matching, knowledge about the joint curvature and conformality index distribution may be used to enhance the orientation field of bad quality latents and infer on the orientation field at bad quality or unobserved locations, e.g \cite{HuckemannHotzMunk2008, BartunekNilssonSallbergClaesson2013}. Further potential applications lie in aiding matching and alignment of fingerprints. In conclusion we remark that correlation between curvature and non-conformality may add essential understanding to embryonic growth models, e.g. \cite{KueckenChampod2013}, which are to date not fully satisfactory. 

\section{Conformal Maps and Quadratic Differentials}

The following can be found in any standard textbook on complex analysis, e.g. \cite{Ahlfors1966}, and on quadratic differentials, e.g. \cite{Strebel1984}.
A complex mapping $x+iy = z\mapsto f(z)$ is conformal if it is partially differentiable with
$$ f'(z)=\frac{\partial f}{\partial  z} = \frac{1}{2}\left(\frac{\partial f(x+iy)}{\partial x} -i\frac{\partial f(x+iy)}{\partial y}\right)\neq 0$$
and if it satisfies the Cauchy-Riemann differential equations
$$\frac{\partial f}{\partial \bar z} = \frac{1}{2}\left(\frac{\partial f(x+iy)}{\partial x} +i\frac{\partial f(x+iy)}{\partial y}\right)= 0\,.$$

\begin{figure}

 \includegraphics[trim=5.5cm 6cm 4.5cm 6cm, clip=true, width=0.4\textwidth]{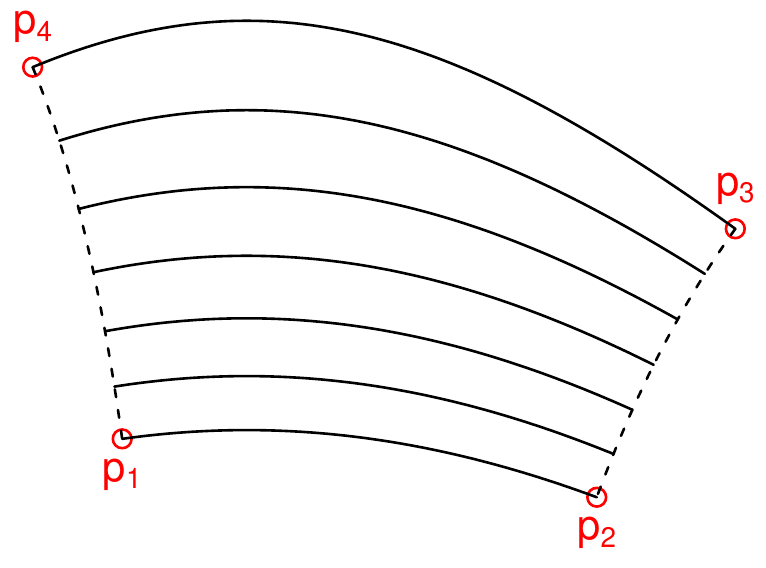} 
 \hspace*{0.1\textwidth}
 \includegraphics[trim=5cm 5cm 5cm 5cm, width=0.35\textwidth]{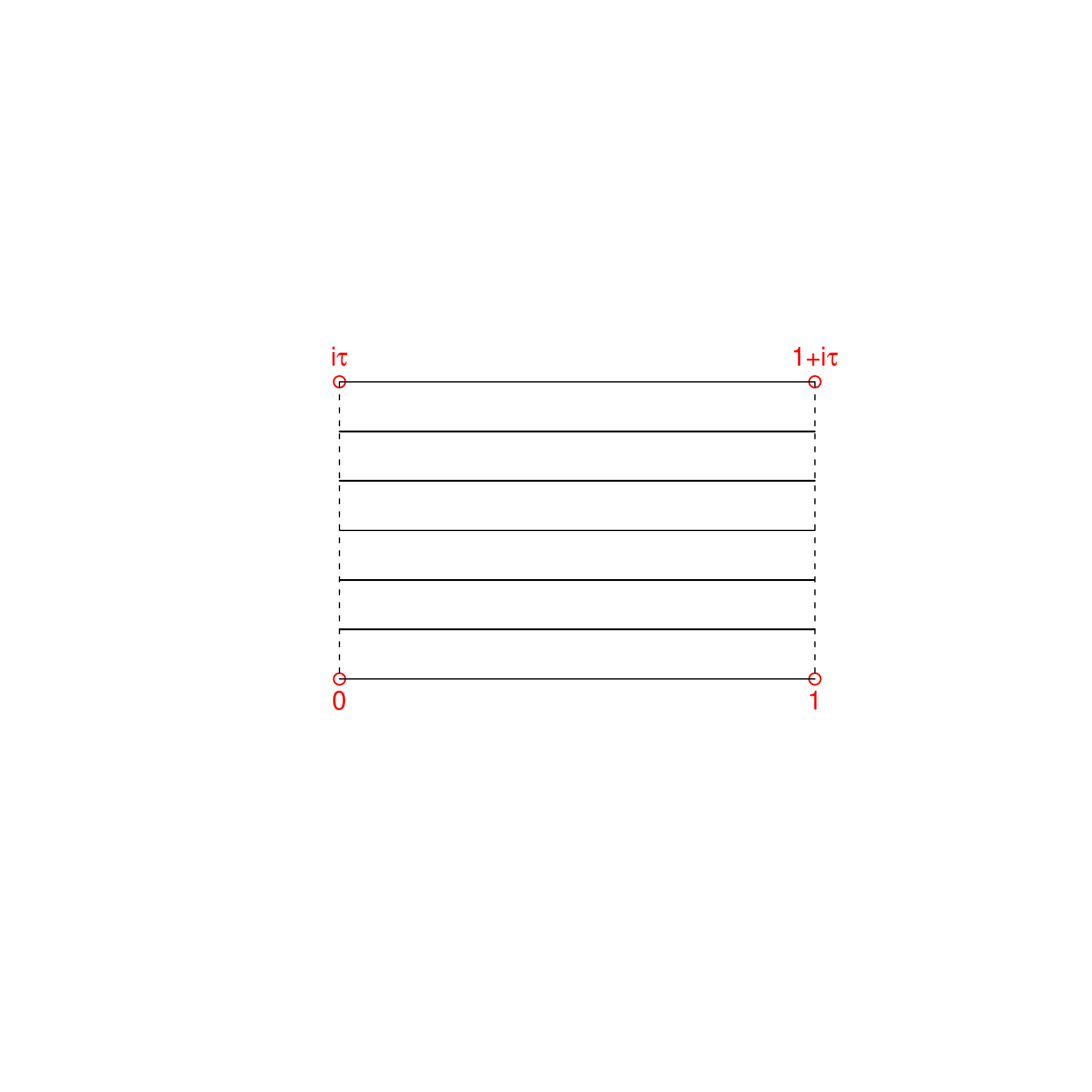}
 
     \hspace*{0.2\textwidth}$z$   \hspace*{0.15\textwidth}       $\longrightarrow$ \hspace*{0.15\textwidth} $w$
 \caption{\it Conformally mapping a quadrilateral with conformal OF from the $z$ plane (left) onto a rectangle in the $w$ plane (right). 
 Trajectories of $dw^2>0$ and their pre-images are depicted by solid lines, orthogonal trajectories by dashed lines.\label{Riemann-mapping-thm:fig}}
\end{figure}

\paragraph{The Riemann Mapping Theorem} asserts that every simply connected open set $G$ that is a proper subset of the complex plane $\mathbb C$ can be mapped conformally onto any open rectangle 
$R_\tau=\{w\in \mathbb C: 0< \RE(w) < 1,~0< \IM(w)<\tau\}$. More precisely, for every selection of four distinct points $p_1,\ldots,p_4 \in \partial G$ on the boundary of $G$  in positive cyclic positive order, there is a unique $\tau>0$ and a unique conformal map $f:G\to R_\tau$ such that
\begin{eqnarray}\label{boundary-correspondence} &p_1 \mapsto 0, p_2 \mapsto 1, p_3\mapsto 1+i\tau\mbox{ and }p_4\mapsto i\tau\,.\end{eqnarray}
The simply connected open set $G$ with these four points is a \emph{quadrilateral} and $\tau=\tau(G)$ is its \emph{modulus}, cf.  Figures \ref{Riemann-mapping-thm:fig} and \ref{deviation-conformality:fig}.
In order to obtain the 
conformal mapping $f$ we introduce quadratic differentials.

\paragraph{A quadratic differential} (QD) $\sigma$ on a subset $D$ of the Riemann sphere 
$\hat{\mathbb C} = \mathbb C \cup \{\infty\}$ is a mapping from the tangent bundle $TD$ into $\hat \CC$ such that 
$$ \sigma(z,dz) = Q(z)\,dz^2$$
with a function $Q$ meromorphic  in $D$, i.e. complex analytic except for possible isolated poles. A \emph{trajectory} of $\sigma$ is a curve $t\mapsto \gamma(t)$ on which $\sigma\big(\gamma(t),\dot\gamma(t)\big)>0$, on an \emph{orthogonal trajectory}, $\sigma <0$. 

If $G$ is a quadrilateral defined by four boundary points $p_1,\ldots,p_4$ with modulus $\tau$, and if $\sigma(z,dz)=Q(z)\,dz^2$ is a quadratic differential, with $Q$ holomorphic in $G$, and trajectories on the boundary of $G$ between $p_1$ and $p_2$ as well as between $p_3$ and $p_4$, and with orthogonal trajectories on the boundary of $G$ between $p_2$ and $p_3$ as well as between $p_4$ and $p_0$, then the above conformal map $f:G \to R_\tau$ with boundary correspondence (\ref{boundary-correspondence}) is given by 
\begin{eqnarray}\label{mapping-QD:eq} f(z) &=& \int_{p_1}^z \sqrt{Q(\zeta)}\,d\zeta\,,\end{eqnarray}
with a suitable branch of the root, cf. Figure \ref{Riemann-mapping-thm:fig}. This results from the differential equation
$$ dw^2 = \big(d f(z)\big)^2 = Q(z)\,dz^2$$
and the fact that all vertical lines are trajectories of $dw^2$, and that horizontal lines are orthogonal trajectories, cf. Figure \ref{Riemann-mapping-thm:fig}.

%

      \paragraph{Circles and M\"obius transformations.} A specific class of conformal mappings is given by  \emph{M\"obius transformation} which have the form
      \begin{eqnarray}\label{Moebius:eq} z\mapsto f(z) &=& \frac{az+b}{cz+d}\,,\end{eqnarray}
      with suitable parameters $a,b,c,d\in \mathbb C$ satisfying $ad-bc \neq 0$. These map $\hat{\mathbb C}$ conformally onto itself so that in particular, circles are mapped to circles. Here, 
      a \emph{circle} in $\hat{\mathbb C}$ is either a proper circle in $\mathbb C$ or a straight line which is then viewed as a circle through $z=\infty$. 
      
      In this terminology, the trajectories of $dw^2$ are then circles passing parallel through $\infty$. Also the orthogonal trajectories of $dw^2$ are circles passing parallel through $\infty$ but orthogonal to the trajectories.
%

\section{Deviation from Conformality}

      \paragraph{Definition.}
      Suppose that we are given an orientation field (OF), arising from a fingerprint, say, in a simply connected domain $G$. This means that its singular points are isolated in $G$ and every non-singular point $z\in G$ carries a unique orientation 
      $ dz^2 \in \mathbb C\setminus\{0\}$. A \emph{trajectory} of the OF is a maximal, differentiable curve $t\mapsto \gamma(t)=z$ through non-singular points such that $(\dot\gamma(t)/dz)^2 >0$. Similarly, for an \emph{orthogonal trajectory} we have $(\dot\gamma(t)/dz)^2 <0$. 
      
\begin{figure}
 
 \includegraphics[trim=5.5cm 6cm 4.5cm 6cm, clip=true, width=0.4\textwidth]{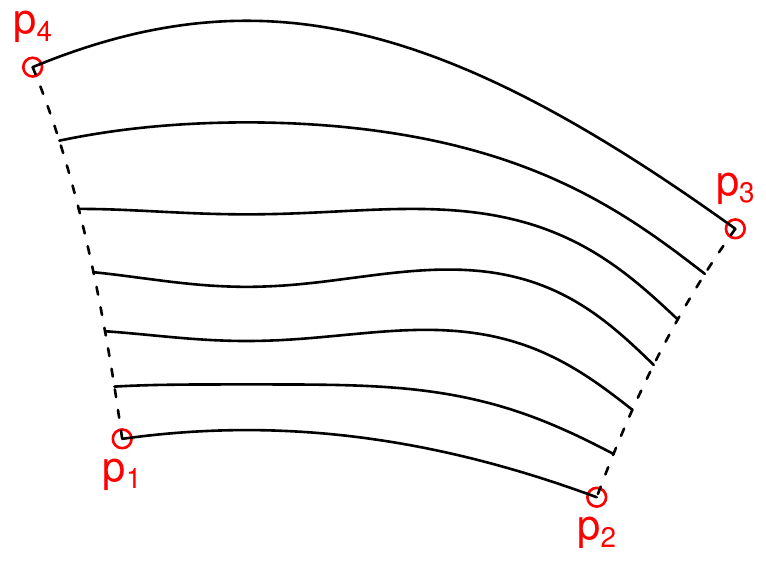} 
 \hspace*{0.1\textwidth}
 \includegraphics[trim=5cm 5cm 5cm 5cm, width=0.35\textwidth]{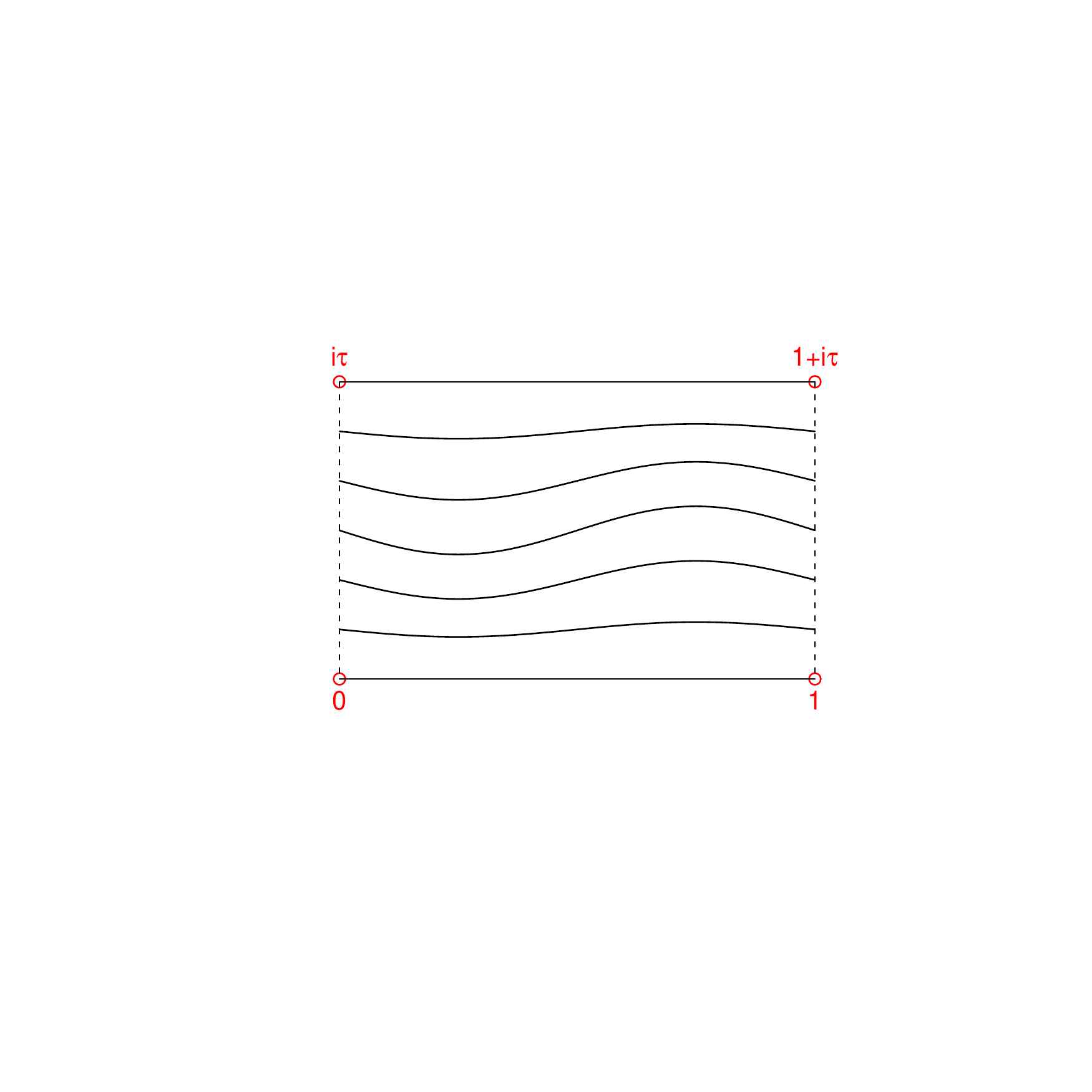}
 
     \hspace*{0.2\textwidth}$z$   \hspace*{0.15\textwidth}       $\longrightarrow$ \hspace*{0.15\textwidth} $w$
 \caption{\it Mapping a quadrilateral with a non-conformal OF conformally onto a rectangle.\label{deviation-conformality:fig}}
\end{figure}

      For such OFs in $G$ we say that a quadrilateral $G_{p_1,p_2,p_3,p_4}\subset G $ is a \emph{subquadrilateral} if 
      \begin{itemize}
       \item its closure comprises non-singular points only,
       \item its boundary comprises arcs of trajectories between $p_1,p_2$ as well as between $p_3,p_4$ and arcs on orthogonal trajectories between $p_2,p_3$ as well as between $p_4,p_1$,
       \item $p_1,p_2,p_3,p_4$ are in cyclic positive order w.r.t. $G_{p_1,p_2,p_3,p_4}$. 
      \end{itemize}
      Under the mapping $f$ from (\ref{mapping-QD:eq}) this subquadrilateral is mapped onto a rectangle with suitable modulus $\tau$. As elaborated in the preceding section, if the OF stems from a QD, under $f$ all trajectories of the OF are mapped to horizontal lines. If this is not the case, we say that the OF \emph{deviates from conformality}, cf. Figure  \ref{deviation-conformality:fig}.
      
      \paragraph{Measuring.}
      In order to measure deviation from conformality several methods come to mind. One could measure, say, in the $w$-plane, 
      \begin{itemize}
       \item maximal vertical aberration of trajectory images from horizontal lines,
       \item or maximal curvature of  trajectory images.
      \end{itemize}
      These and similar method require in particular the numerical computation of the integral in (\ref{mapping-QD:eq}), a numerically highly challenging problem. 
      
      In a first approximation, one might replace the trajectory arcs of the OF with circular arc segments giving a \emph{circular arc rectangle}. 
      Then the integral in (\ref{mapping-QD:eq}) could be analytically solved by taking recourse to the circular version of the \emph{Schwarz-Christoffel Formula}. This formula, however, as is well known, is numerically highly unstable, cf. \cite{DriscollTrefethen2002}.
      
\begin{figure}[t!]   
      \centering
      \includegraphics[width=0.9\textwidth]{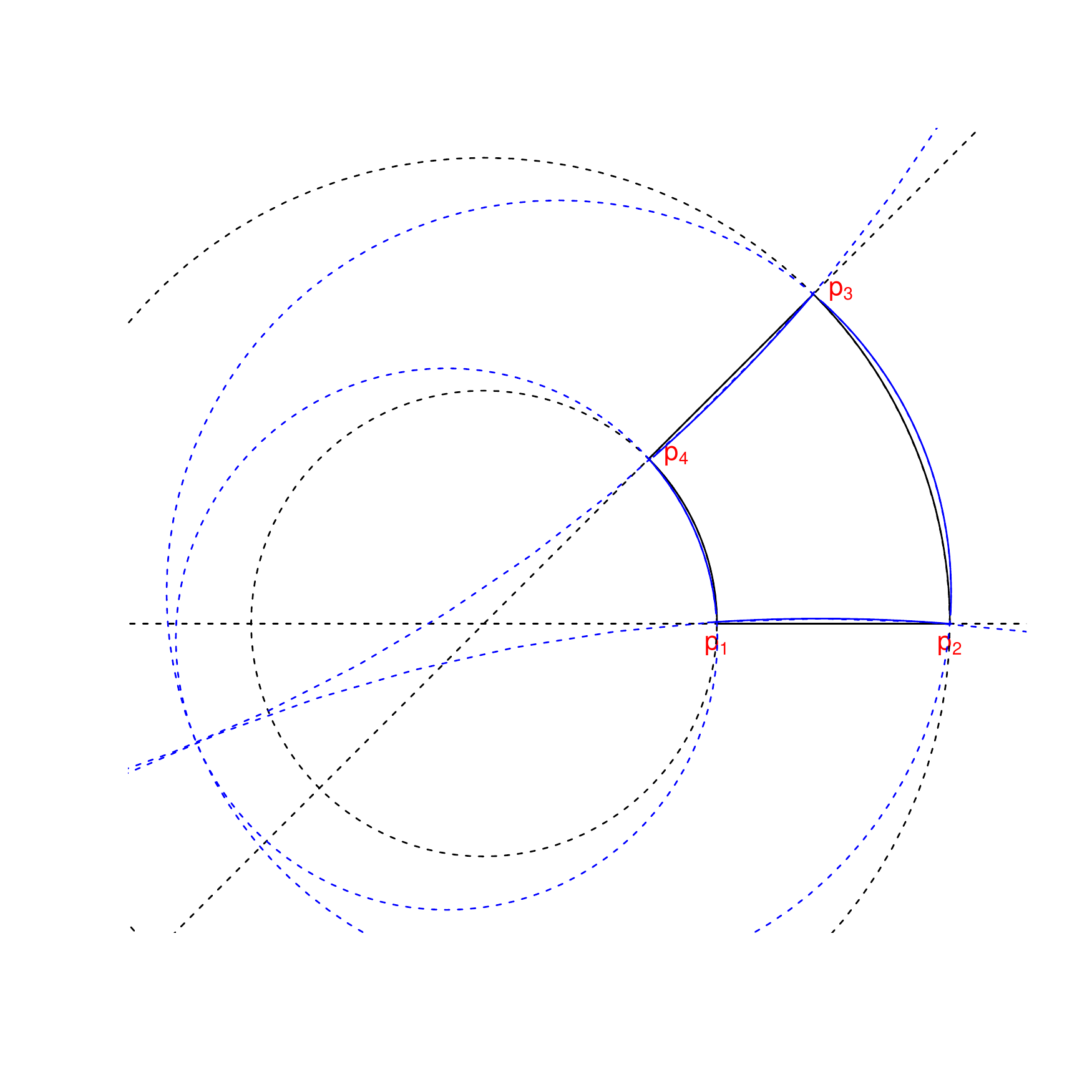}
       \caption{\it A circular arc rectangle (black) formed by two intersecting lines and two concentric circles and its M\"obius approximation (blue) formed by two pairs of circles, each pair is tangential at a common point and intersecting the other pair orthogonally there. Both circular arc rectangles have the same vertex points $p_1,p_2,p_3,p_4$ where the corresponding circles intersect orthogonally.
			          \label{SCF-circular-Moebius:fig}}
			

      \end{figure}

      \paragraph{The M\"obius approximation.} In a second approximation we replace the trajectory and orthogonal trajectory arcs on the boundary of $G_{p_1,p_2,p_3,p_4}$ 
      with specific circular arc segments, mimicking the trajectory and orthogonal trajectory structure of $dw^2$ as discussed above, such that the corresponding circles intersect at a common point in $\hat{\mathbb C}$ where the circles determined by $p_1,p_2$ and $p_3,p_4$ are parallel and orthogonal to the circles determined by $p_2,p_3$ and $p_4,p_1$, which are also parallel there. Then 
      the conformal mapping $f$ from (\ref{mapping-QD:eq}) is simply a M\"obius transformation 
      given by (\ref{Moebius:eq}).
      In this case the modulus of $G_{p_1,p_2,p_3,p_4}$ is given by the double cross-ratio
      \begin{eqnarray}\label{modulus:eq}
     \tau(G_{p_1,p_2,p_3,p_4}) &=& \frac{p_1-p_2}{p_1-p_4}~\frac{p_3-p_4}{p_3-p_2}\,. 
      \end{eqnarray}
      For general subquadrilaterals $G_{p_1,p_2,p_3,p_4}$ denote the right hand side of (\ref{modulus:eq}) by
      \begin{eqnarray}\label{modulus2:eq}
            M(G_{p_1,p_2,p_3,p_4}) &:=& \frac{p_1-p_2}{p_1-p_4}~\frac{p_3-p_4}{p_3-p_2}\,.
            \end{eqnarray}
      Figure \ref{SCF-circular-Moebius:fig} shows the general case of a circular arc rectangle and its M\"obius approximation.
      
\begin{figure}
      \centering
      \includegraphics[width=0.5\textwidth]{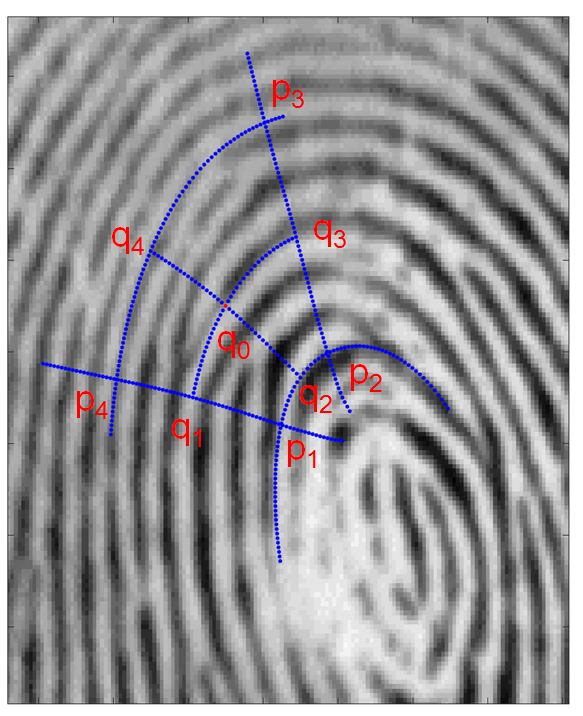}
       \caption{\it A tetraquadrilateral obtained from choosing the point $q_0$ within a subquadrilateral.\label{TQL:fig}}  
      \end{figure}

      \paragraph{Tetraquadrilaterals and M\"obius moduli.} 
      In order to propose a simple measure for the degree of conformality, we subdivide a given subquadrilateral $G_{p_1,p_2,p_3,p_4}$ of an OF into four smaller subquadrilaterals, determined by picking an arbitrary but fixed point $q_0 \in G_{p_1,p_2,p_3,p_4}$. Then the OF's trajectory through $q_0$ intersects the orthogonal trajectories between $p_1,p_4$ in a point $q_1$ and between $p_2,p_3$ in a point $q_3$, respectively and the OF's orthogonal  trajectory through $q_0$ intersects the trajectories between $p_1,p_2$ in a point $q_2$ and between $p_3,p_4$ in a point $q_4$, respectively, cf. Figure \ref{TQL:fig}. We call this subdivision into four subquadrilaterals a \emph{tetraquadrilateral} (TQL) centered at $q_0$. We apply the M\"obius approximation jointly to each of the four subquadrilaterals, i.e. we assume that all trajectories are on circles that touch at a common point in $\hat{\mathbb C}$ with tangent orthogonal to the tangent of the circles on which the orthogonal trajectories lie, that touch at the same point. Then the cross-ratio of moduli -- we take the approximation $M$ given by (\ref{modulus2:eq}) -- gives rise to the \emph{M\"obius modulus}  
      \begin{eqnarray}\nonumber\label{Moebius-modulus:eq}
       M(G_{p_1,p_2,p_3,p_4},q_0)&=& \frac{ M(G_{p_1,q_2,q_0,q_1})}{M(G_{q_2,p_2,q_3,q_0})}~\frac{M(G_{q_0,q_3,p_3,q_4})}{M(G_{q_1,q_0,q_4,p_4})}\\
       &=& \frac{p_1-q_2}{q_2-p_2}~\frac{p_2-q_3}{q_3-p_3}~\frac{p_3-q_4}{q_4-p_4}~\frac{p_4-q_1}{q_1-p_1}\,.
      \end{eqnarray}
      This is a complex number that only vanishes in degenerate scenarios which we have excluded by definition. If the M\"obius approximation holds true and if the OF is conformal, then $M(G_{p_1,p_2,p_3,p_4},q_0)=1$. Otherwise, $\log M$ gives in approximation the deviation from conformality.
    
	Simulations with OFs stemming from QDs show that the M\"obius approximation is quite good. In fact, below it serves surprisingly well to identify areas of deviation from conformality of real and synthetic fingerprints.

\begin{figure}
            \centering
      \includegraphics[width=0.45\textwidth]{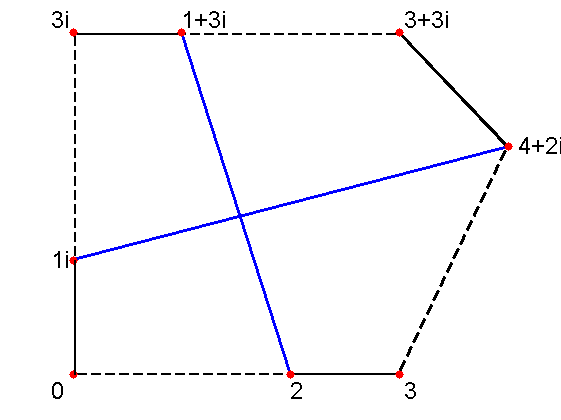}
      \includegraphics[width=0.45\textwidth]{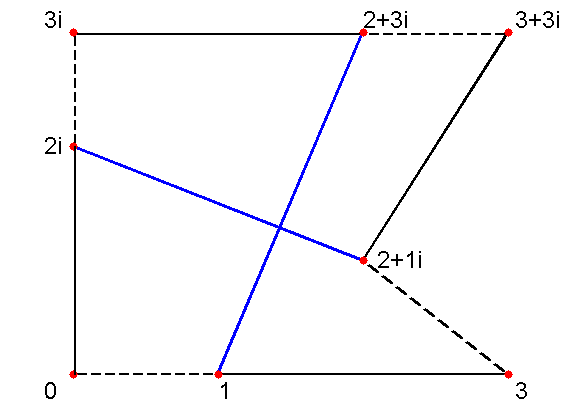}
       \caption{\it Visualizing the Moebius modulus $M=me^{i\phi}$ of a TQL. Under conformality $M=1$, if the M\"obius approximation is exact. Left: Positive $\log m$ and $\phi \in (-\pi,0)$. Right: Negative $\log m$ and $\phi \in (0,\pi)$.\label{Visualizing-Moebius-moduli:fig}}  
      \end{figure}

	\paragraph{Visualizing M\"obius moduli of tetraquadrilaterals.} From equation (\ref{Moebius-modulus:eq}) we see that the M\"obius modulus $M = m e^{i\phi}$ of a TQL is determined by the ratios of subsequent boundary segments. The logarithm of its absolute value, $\log m$,  called the \emph{conformality index}, reflects the ratio of the lengths and its argument $\phi$ reflects the sum and differences of the angles, cf. Figure \ref{Visualizing-Moebius-moduli:fig}. In the application of this contribution we only consider the conformality index.

\section{Estimating M\"obius Moduli in Fingerprints} \label{sec:EstimatingMoebiusModuli}

\begin{figure}    
      \centering
      \includegraphics[width=0.5\textwidth]{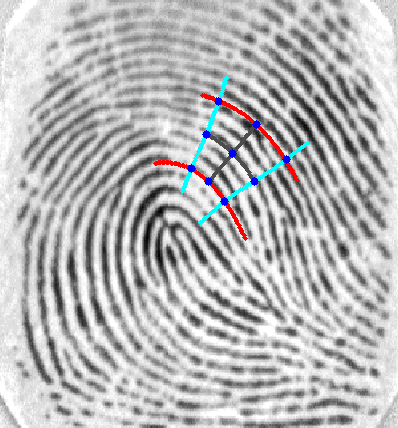}
       \caption{\it Stepwise building a TQL: First an equal sided cross aligned with the OF about the central point (grey), then traversing along trajectories (red) and orthogonal trajectories (light blue), respectively, from the endpoints, until intersection points are reached. These intersection points, with the crosses endpoints and the central point give the TQL (dark blue). 
			          \label{Building-TQL:fig}}  
\end{figure}

Fingerprint segmentation \cite{ThaiHuckemannGottschlich2016,ThaiGottschlich2016G3PD}
into foreground, the \emph{region of interest} (ROI), and background, as well as orientation field estimation by a combination of the line sensor \cite{GottschlichMihailescuMunk2009} and gradient based method
as described in \cite{GottschlichSchoenlieb2012}, say, are the first two typical processing steps in fingerprint algorithms. For a given fingerprint image with estimated ROI  and orientation field,  
%
%
%
%
%
%
in order to define TQLs, we use the \emph{curved regions} of \cite{Gottschlich2012}: A TQL centered at $q_0$ in the ROI can be obtained as follows. From $q_0$ traverse the OF in either direction with a fixed length $c=40$ pixels (approximately 3 ridge distances) and the orthogonal OF also in either direction with same lengths $c$. This gives the black cross in Figure \ref{Building-TQL:fig}. From the endpoints of that cross, traverse the OF in either direction (red in Figure \ref{Building-TQL:fig}), or the orthogonal OF in either direction (light blue in Figure \ref{Building-TQL:fig}), respectively, until the corresponding curves meet. The four intersection points together with the four crosses' endpoints and $q_0$ form the TQL. We compute TQLs centered at every pixel locus $q_0$ over the entire ROI.


\begin{figure}
            \centering
      \includegraphics[width=0.48\textwidth]{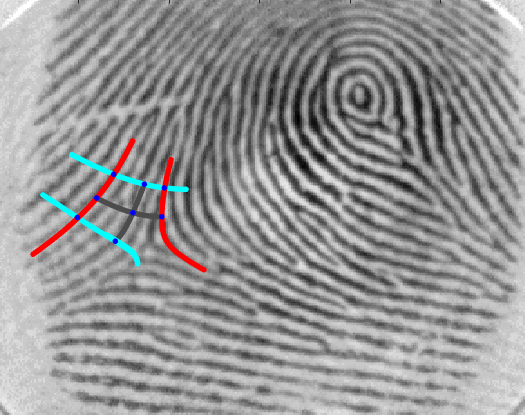}
      \includegraphics[width=0.48\textwidth]{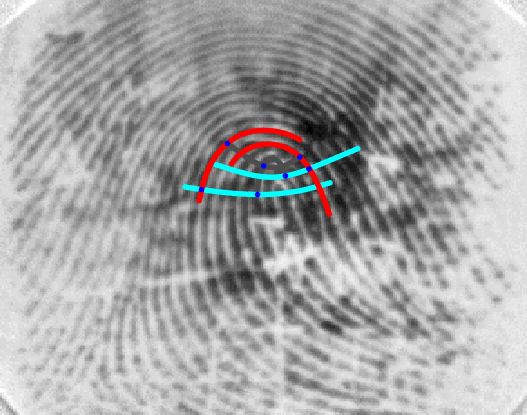}
       \caption{\it Building a TQL close to a singular point may fail if trajectories turn away from one another, near a delta (left), or self intersect, near a core (right).\label{Singularities:fig}}  
      \end{figure}

If in the process of the above routine, trajectories or orthogonal trajectories leave the ROI, no TQL is computed. Moreover, two problems can occur, when attempting to determine TQLs close to singular points, and also in these cases, no TQLs are computed. First, it may happen that orthogonal and non-orthogonal trajectories turn away from one another, and hence, have no intersection points, as depicted in Figure \ref{Singularities:fig} (left panel). Secondly, high curvature can lead to self intersecting lines, say, of the central cross in Figure \ref{Singularities:fig} (right panel), so that no meaningful TQL can be defined.

%
%
%
%

\section{Synthetic Fingerprint Generation}

The generation of synthetic fingerprints is of great interest to the biometric and forensic community.
Fingerprint databases are required for 
evaluating and comparing the performance of algorithms 
for minutiae extraction, fingerprint verification,
fingerprint indexing and identification.
In the following, we focus on artificially generated fingerprint images.
A very related but different topic is fingerprint liveness detection,
the discrimination between images of real, alive fingers and images of spoof fingers 
made from material like gelatin, wood glue or silicone (see \cite{Gottschlich2016} for a recent survey on fingerprint liveness detection).

Major advantages of synthetic fingerprints are given by the fact that millions of prints can be created at virtually no cost
and their generation is not hindered by national laws and legal constraints concerning data protection and privacy.
However, artificial fingerprints have to be 'realistic',
or otherwise the validity of results obtained on databases of synthetic prints can be called into question.
In 2014, a study showed that the methods for fingerprint generation at that time 
produced prints with an unrealistic minutiae distributions \cite{GottschlichHuckemann2014}:
Distances between minutiae locations and angles between minutiae directions,
summarized in minutiae histograms (MHs) \cite{GottschlichHuckemann2014}
for minutiae pairs had been compared by the earth movers' distance (EMD) \cite{GottschlichSchuhmacher2014}.

Methods for creating artificial fingerprint include 
SFinGe by Cappelli \textit{et al.} \cite{CappelliErolMaioMaltoni2000}
and Araque \textit{et al.} \cite{AraqueEtAl2002}. 
Both algorithms utilize a global orientation field model by 
Vizcaya and Gerhardt \cite{VizcayaGerhardt1996}.
Images created by SFinGe are part of the widely used FVC databases.

While previously minutiae have been used to separate real
from synthetic fingerprints \cite{GottschlichHuckemann2014},
the experiments and results described in the next section 
are based on the orientation field only.

SFinGe has also been used in comparison of fingerprint classification methods
by Galar \textit{et al.} \cite{GalarClassification1,GalarClassification2}.
Most classification methods utilize the orientation field, and hence, 
these results are based on unrealistic OFs as will be shown in the next section.

Recently, a novel method for fingerprint generation has been proposed \cite{ImdahlHuckemannGottschlich2015}
which overcomes the aforementioned problems.
The realistic fingerprint creator (RFC) \cite{ImdahlHuckemannGottschlich2015}
utilizes orientation fields from real fingerprints 
and templates are checked whether they pass the 'test of realness' \cite{GottschlichHuckemann2014}
for minutiae distribution.

\section{Discriminating Real from Synthetic Prints by Histograms of M\"obius Moduli and Curvatures} \label{sec:results}


M\"obius moduli and curvature \cite{Gottschlich2012} are computed as described in Section \ref{sec:EstimatingMoebiusModuli}. Then
a 2D histogram summarizes the joint distribution of curvature and conformality indices (logs of absolute values of M\"obius moduli) for every foreground pixel for which a TQL was computed.

These 2D histograms with 10 bins, and with 20 bins, for each dimension, 
are considered as feature vectors with 100 and 400 entries, respectively, 
and these feature vectors are used for training
by a support vector machine with a linear kernel ($C = 1.0$).
Experimental results listed in Table \ref{tab:results}
have been obtained using the software package LIBSVM \cite{ChangLin2011}.

Each FVC database contains images from 110 fingers with 8 impressions per finger.
Each competition in 2000, 2002 and 2004 contains one synthetic database (DB 4) 
which we have paired with one real database (DB 1) from the same year.
From 110 available fingers,
60 real and 60 synthetic are assigned into the training set
and the remaining 50 real and 50 synthetic into the test set.

\begin{table}[h!]
	\centering
\begin{tabular}{|l|ccc|ccc|}
\hline
 & \multicolumn{3}{c|}{10 bins, one impression} & \multicolumn{3}{c|}{20 bins, one impression} \\ \hline
 FVC & 2000 & 2002 & 2004 & 2000 & 2002 & 2004  \\  \hline
 Training set	&96.7\%  & 98.3\% & 97.5\% & 100\% & 100\%& 100\% \\
 Test	set &76\% &	73\% & 70\% & 74\% & 80\% & 75\% \\  \hline
 & \multicolumn{3}{c|}{10 bins, eight impressions} & \multicolumn{3}{c|}{20 bins, eight impressions} \\ \hline
 FVC & 2000 & 2002 & 2004 & 2000 & 2002 & 2004  \\  \hline
 Training	& 100\% & 100\% & 100\% & 100\% & 100\% & 100\%  \\
 Test	set & 100\% & 100\% & 100\% & 100\% & 100\% & 100\%  \\
 \hline
\end{tabular}
\caption{\it SVM classification accuracy of 2D histograms using 10 (left) and 20 bins (right) 
for each of the two dimensions, M\"obius moduli and curvature. 
The results in the upper half summarize histograms per one impression, in the lower half, histograms per eight impressions.
\label{tab:results} } 
\end{table}

\begin{figure}[t!]
\centering
\subfigure{
	\includegraphics[width=0.3\textwidth]{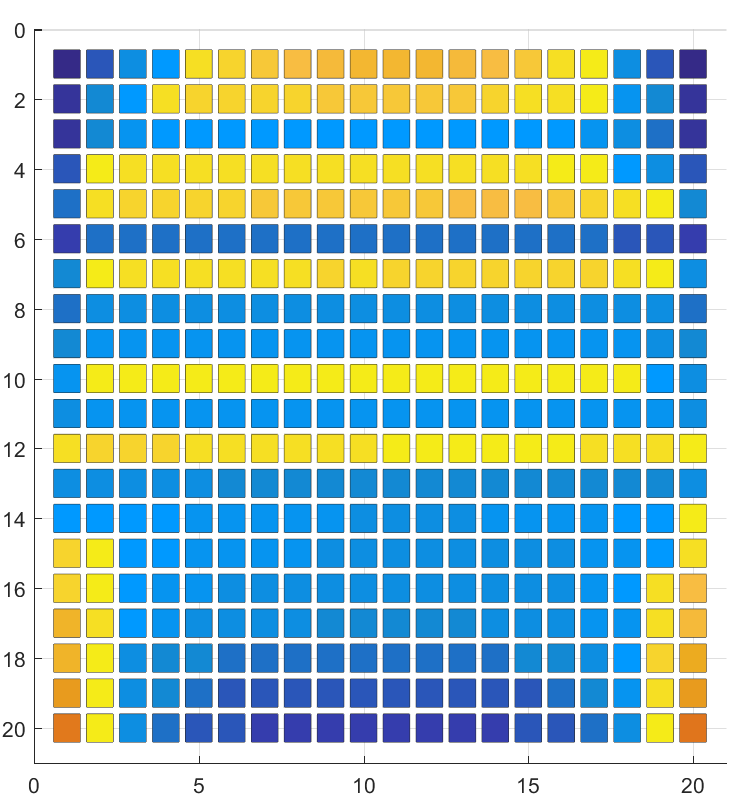}}
\subfigure{
	\includegraphics[width=0.3\textwidth]{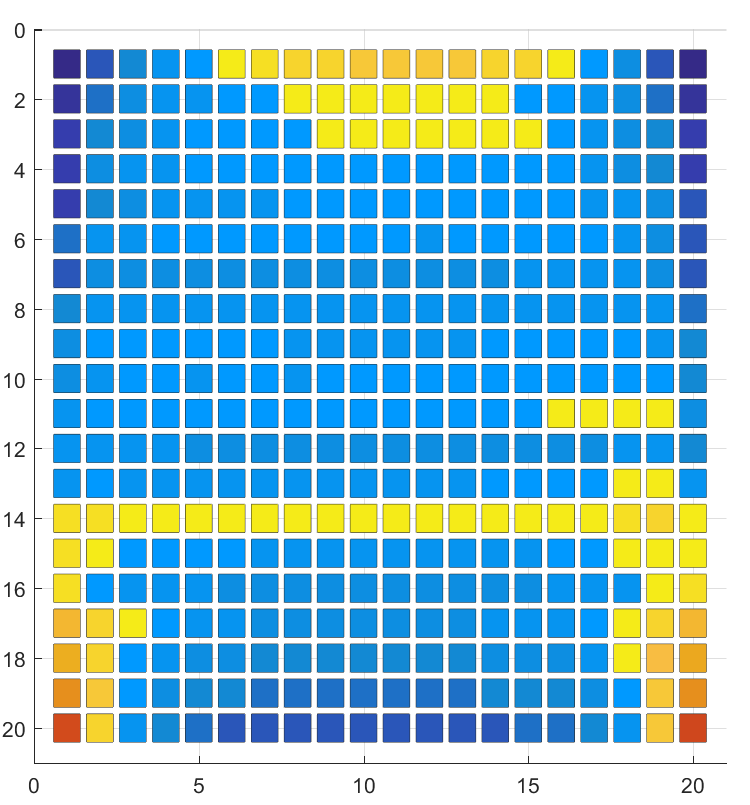}}
\subfigure{
	\includegraphics[width=0.3\textwidth]{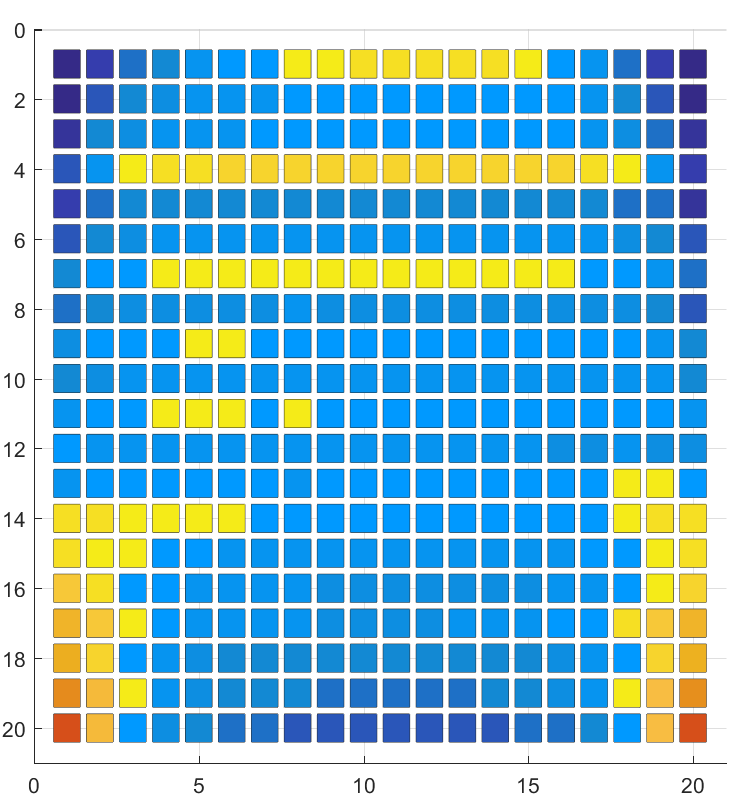}}
\subfigure{
	\includegraphics[width=0.3\textwidth]{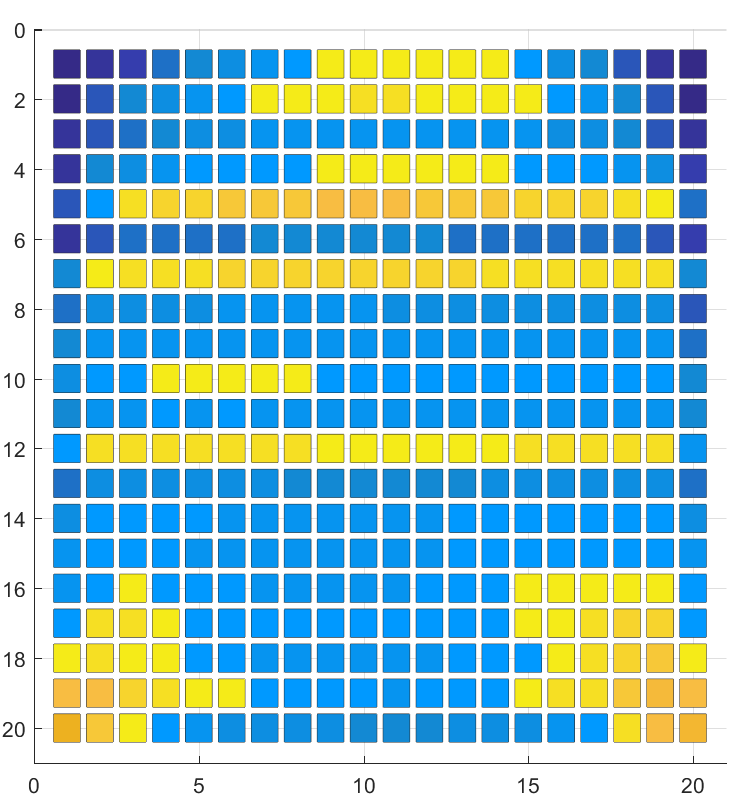}}
\subfigure{
	\includegraphics[width=0.3\textwidth]{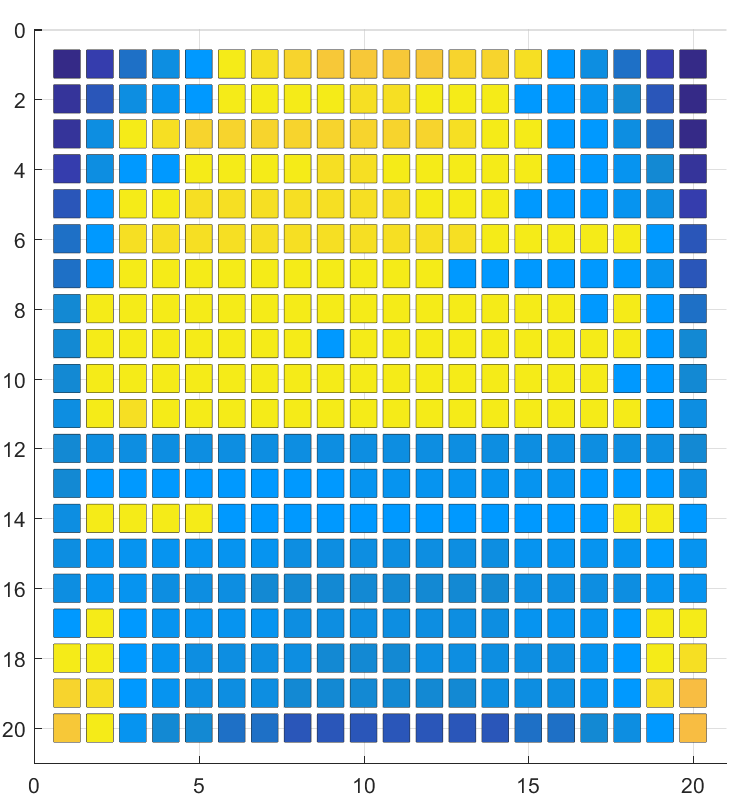}}
\subfigure{
	\includegraphics[width=0.3\textwidth]{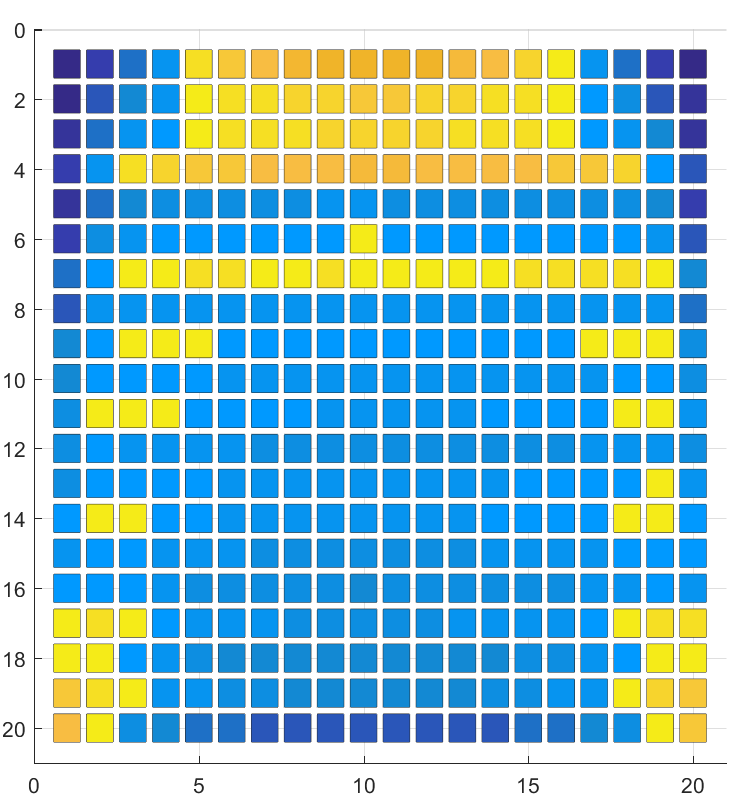}}
\caption{\it Averaged histograms of conformality indices (horizontal, zero in the middle and $\pm 0.33$, and beyond, at the ends) vs. curvature (vertical, zero on top, $2.6$, and beyond, at the bottom) for FVC 2000 (left column), FVC 2002 (middle column) and FVC 2004 (right column). Top row: Real fingerprints.  Bottom row: Synthetic fingerprints.
\label{durchschnitt}}
\end{figure}

\section{Discussion}

\begin{figure}[t!]
\centering
\subfigure{
	\includegraphics[width=0.14\textwidth]{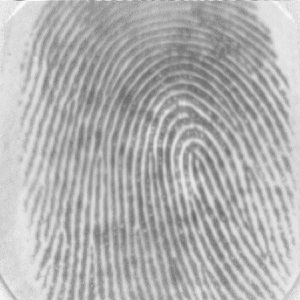}}
\subfigure{
	\includegraphics[width=0.14\textwidth]{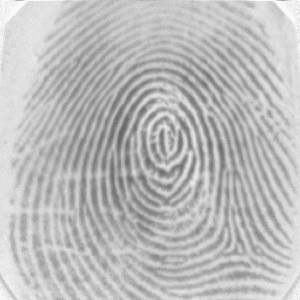}}
\subfigure{
	\includegraphics[width=0.14\textwidth]{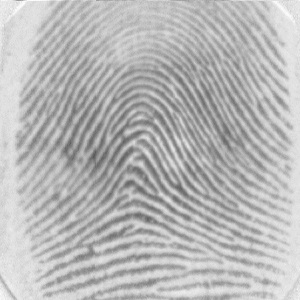}}
\subfigure{
	\includegraphics[width=0.14\textwidth]{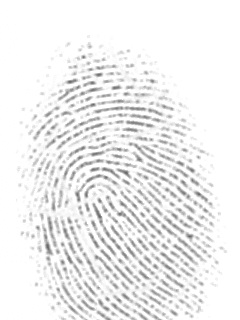}}
\subfigure{
	\includegraphics[width=0.14\textwidth]{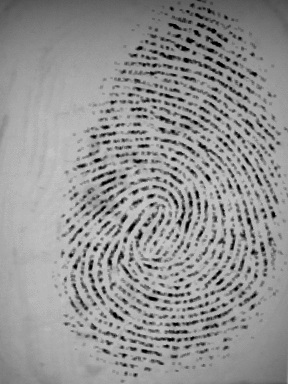}}
\subfigure{
	\includegraphics[width=0.14\textwidth]{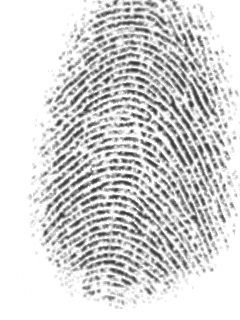}}\\
\subfigure{
	\includegraphics[width=0.14\textwidth]{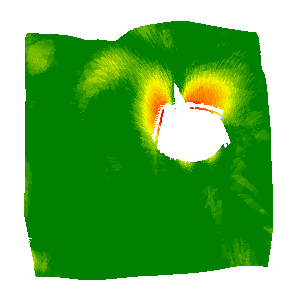}}
\subfigure{
	\includegraphics[width=0.14\textwidth]{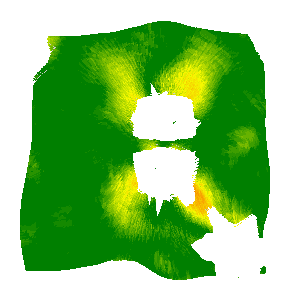}}
\subfigure{
	\includegraphics[width=0.14\textwidth]{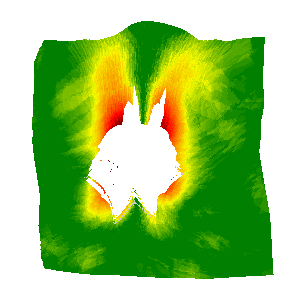}}
\subfigure{
	\includegraphics[width=0.14\textwidth]{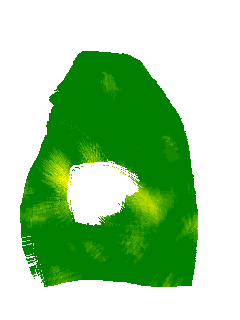}}
\subfigure{
	\includegraphics[width=0.14\textwidth]{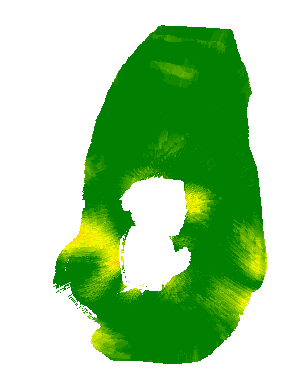}}
\subfigure{
	\includegraphics[width=0.14\textwidth]{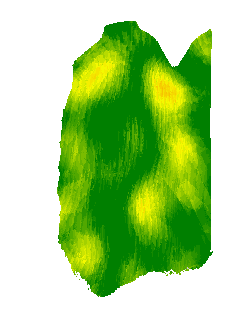}}\\
\subfigure{
	\includegraphics[width=0.14\textwidth]{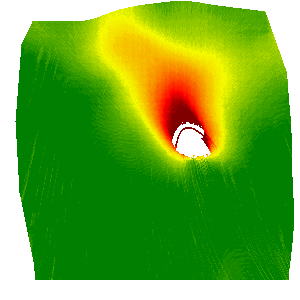}}
\subfigure{
	\includegraphics[width=0.14\textwidth]{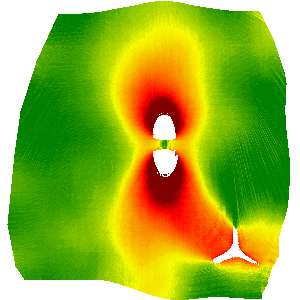}}
\subfigure{
	\includegraphics[width=0.14\textwidth]{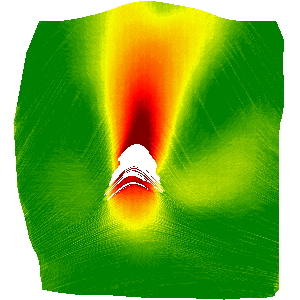}}
\subfigure{
	\includegraphics[width=0.14\textwidth]{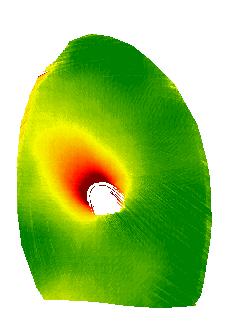}}
\subfigure{
	\includegraphics[width=0.14\textwidth]{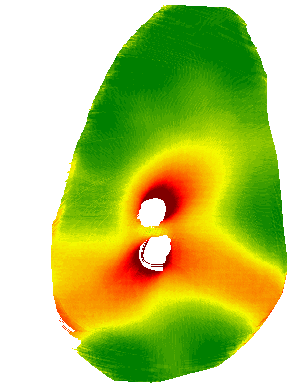}}
\subfigure{
	\includegraphics[width=0.14\textwidth]{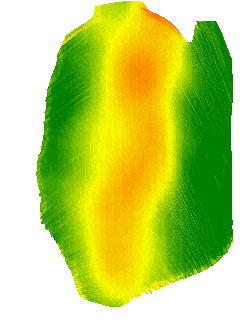}}\\
\subfigure{
	\includegraphics[width=0.14\textwidth]{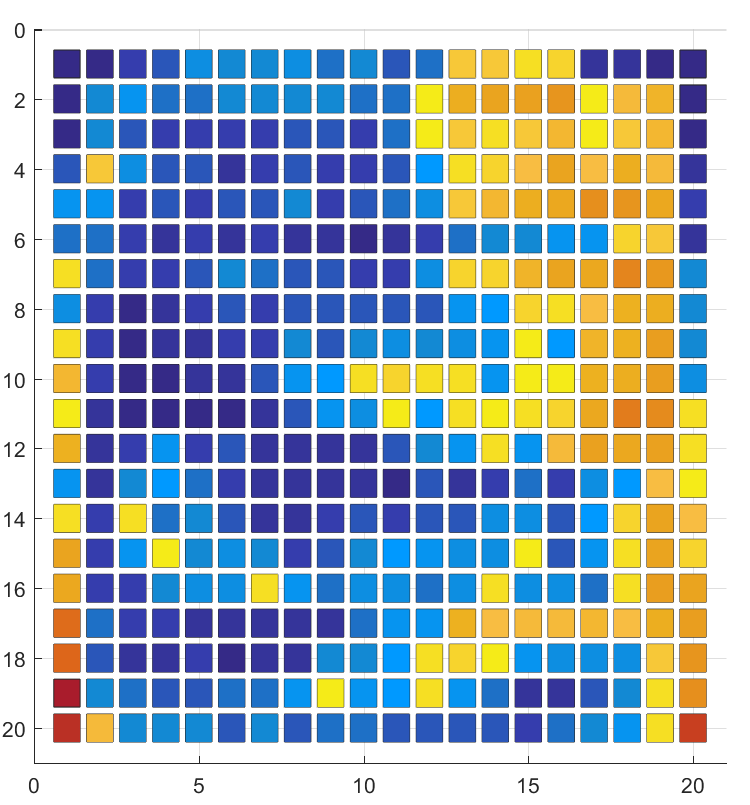}}
\subfigure{
	\includegraphics[width=0.14\textwidth]{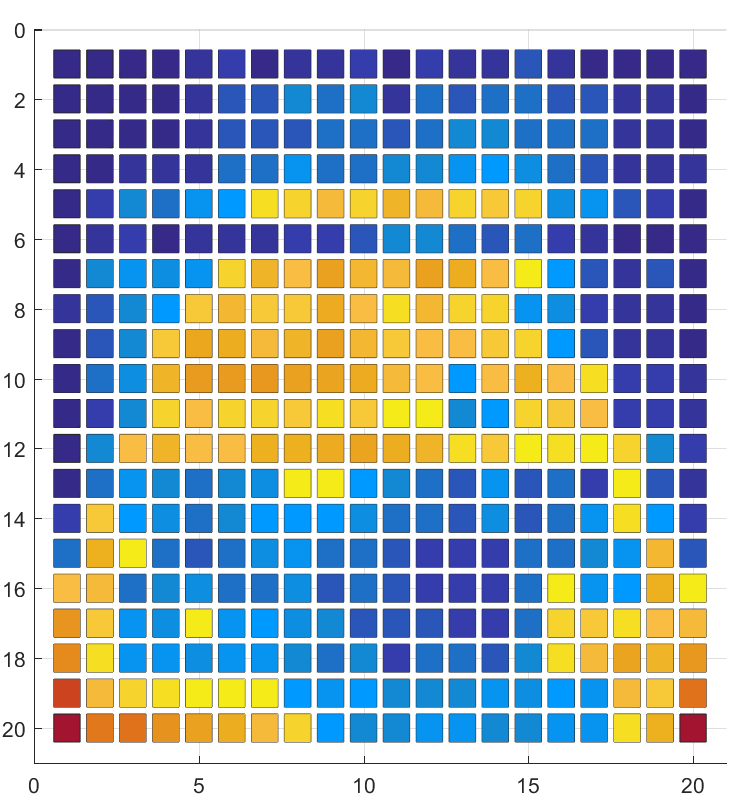}}
\subfigure{
	\includegraphics[width=0.14\textwidth]{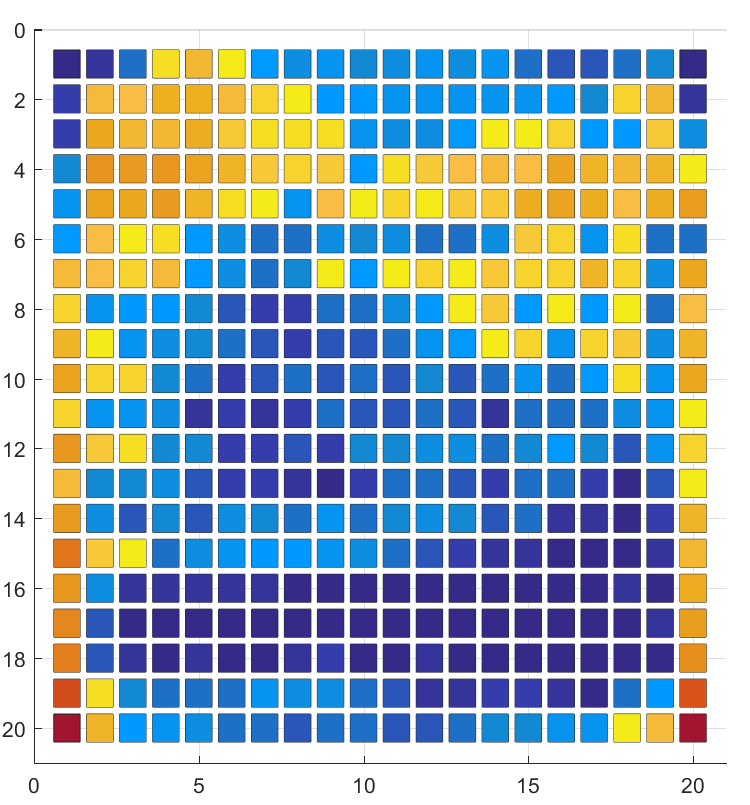}}
\subfigure{
	\includegraphics[width=0.14\textwidth]{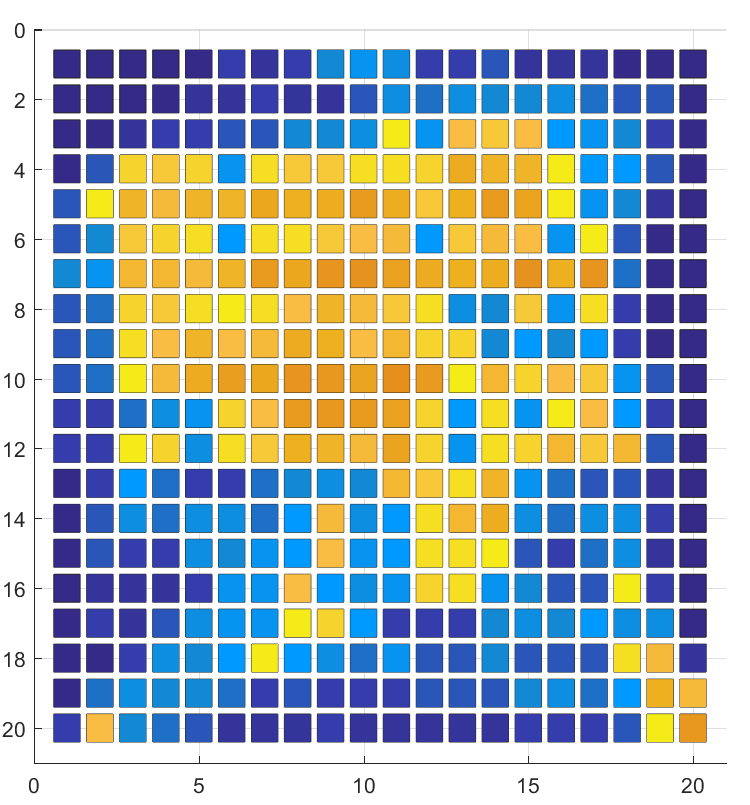}}
\subfigure{
	\includegraphics[width=0.14\textwidth]{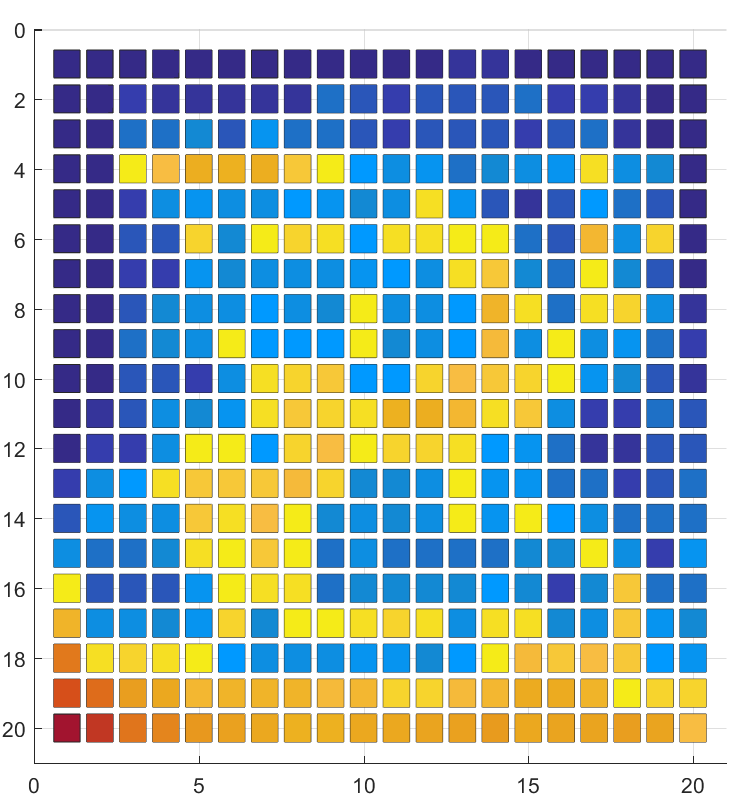}}
\subfigure{
	\includegraphics[width=0.14\textwidth]{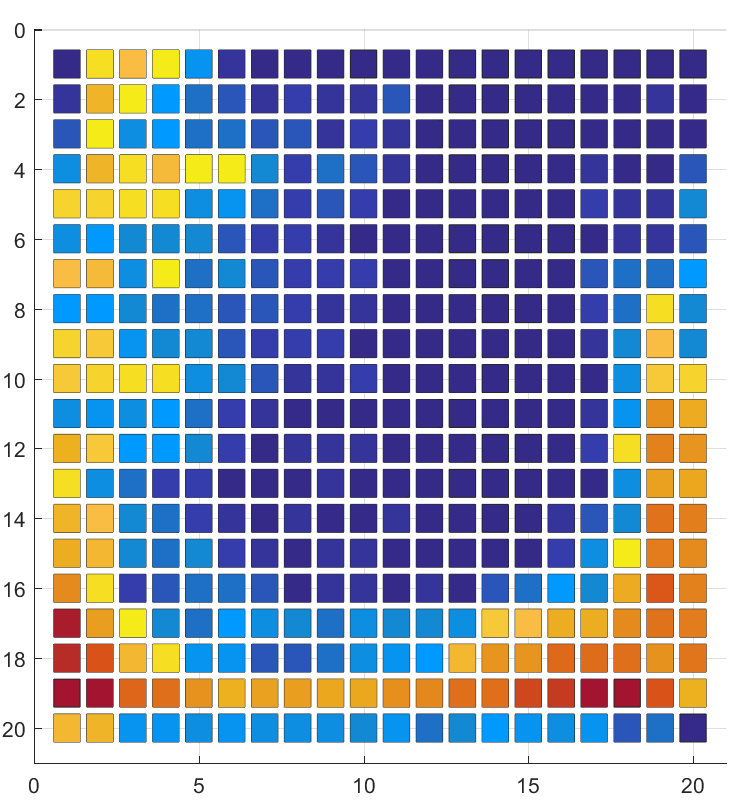}}
\caption{\it Exemplary fingerprints (top row) with conformality index field (2nd row), curvature field (3rd row) and joint histograms (bottom row, axes as in Fig, \ref{durchschnitt}). The left three column are from real fingers, the right three columns from synthetic prints.}\label{finger}
\end{figure}

In Figure \ref{durchschnitt} we see that for real fingerprints (top row), extremal conformality indices  (first and last histogram columns) strongly peak at high curvature (bottom histogram rows), whereas for synthetic prints, these peaks are far less pronounced. These systematic differences are also visible in Figure \ref{finger}. 
Upon closer inspection note that high curvature regions form similar clusters for both types of prints, these clusters, however, split into two extremal non-conformality clusters, only for real prints, like \emph{rabbit ears}. 
These rabbit ears indicate that above cores and above and below whorls, real fingerprints are locally more bent or distorted than as predicted by the corresponding zero-pole model part in a quadratic differential model. In real prints, high curvature and extremal non-conformality occur both on the right and left above cores and above and below whorles. In synthetic prints, high curvature occurs at similar locations, but extremal non-conformality is usually not present there.


Results in Table \ref{tab:results} underline the 
high discriminability which the proposed 2D histogram of curvature and conformality indices 
provides.
Both proposed features are computed using the estimated orientation field of a fingerprint.
This confirms the conclusion that there are systematic differences 
between the orientation fields of real fingerprints 
and those created from models in the biometrics literature
like the Vizcaya-Gerhardt model \cite{VizcayaGerhardt1996} 
used in SFinGe \textit{et al.} \cite{CappelliErolMaioMaltoni2000}.
The proposed feature encodes this systematic difference in a 2D histogram.

Considering the results for 20 bins per dimension 
and histograms based on one impression in Table \ref{tab:results},
we observe classification accuracies on the training set of 100\%
and accuracies on the test set between 74 and 80\%.
We believe that the limiting factor for the accuracy is the very small number 
of training examples (60 real and 60 synthetic images)
which could be resolved by larger databases.

Interestingly, histograms which summarize eight instead of just one impression
achieve a perfect classification accuracy on both training and test set
despite the small number of training examples.
Computing an average histogram 
which captures the average joint distribution of curvature and M\"obius moduli 
over several (eight) impressions seems to be robust 
under fluctuations in histograms between different impressions of the same finger.

In conclusion we remark that new algorithms for generating realistic synthetic orientation fields have to be developed.  
The recently proposed XQD model \cite{GottschlichTamsHuckemann2017} could be used for this purpose, realistically linking curvature with conformality indeces.
Until their arrival, 
one may rely on 
orientation fields from real fingerprints (as implemented by RFC \cite{ImdahlHuckemannGottschlich2015}).

\section*{Acknowledgements}

The first three authors gratefully acknowledge support from the 
Felix-Bernstein-Institute for Mathematical Statistics in the Biosciences, 
the Niedersachsen Vorab of the Volkswagen Foundation and the DFG Graduate Research School 2088. 
The last two authors express their gratitude for support from the HeKKSaGOn cooperation.
Stephan Huckemann also expresses gratitude for support by 
the SAMSI Forensics Program 2015/16.

\bibliographystyle{../../BIB/Chicago}

\bibliography{Moebius}

\end{document}